\documentclass[aps,prl,twocolumn,superscriptaddress,showpacs,floatfix,longbibliography]{revtex4-1}

\makeatletter
\def\@bibdataout@aps{%
\immediate\write\@bibdataout{%
@CONTROL{%
apsrev41Control%
\longbibliography@sw{%
    ,author="08",editor="1",pages="1",title="0",year="1"%
    }{%
    ,author="08",editor="1",pages="1",title="",year="1"%
    }%
  }%
}%
\if@filesw \immediate \write \@auxout {\string \citation {apsrev41Control}}\fi
}
\makeatother

\usepackage{graphicx}
\usepackage{dcolumn}
\usepackage{bm}

\usepackage{multirow}

\usepackage{amssymb}
\usepackage{amsmath}
\usepackage{graphicx}
\usepackage{xcolor}
\usepackage[colorlinks,citecolor=blue,linkcolor=red,urlcolor=blue]{hyperref}

\begin{document}
\title{Relaxation Critical Dynamics in Measurement-induced Phase Transitions}
\author{Wantao Wang}
\affiliation{Guangdong Provincial Key Laboratory of Magnetoelectric Physics and Devices, Sun Yat-Sen University, Guangzhou 510275, China}
\affiliation{School of Physics, Sun Yat-Sen University, Guangzhou 510275, China}
\author{Shuo Liu}
\affiliation{Institute for Advanced Study, Tsinghua University, Beijing 100084, China}
\author{Jiaqiang Li}
\affiliation{Guangdong Provincial Key Laboratory of Magnetoelectric Physics and Devices, Sun Yat-Sen University, Guangzhou 510275, China}
\affiliation{School of Physics, Sun Yat-Sen University, Guangzhou 510275, China}
\author{Shi-Xin Zhang}
\affiliation{Institute of Physics, Chinese Academy of Sciences, Beijing 100190, China}
\author{Shuai Yin}
\email{yinsh6@mail.sysu.edu.cn}
\affiliation{Guangdong Provincial Key Laboratory of Magnetoelectric Physics and Devices, Sun Yat-Sen University, Guangzhou 510275, China}
\affiliation{School of Physics, Sun Yat-Sen University, Guangzhou 510275, China}

\date{\today}

\begin{abstract}
Measurement-induced phase transition (MIPT) describes the nonanalytical change of the entanglement entropy resulting from the interplay between measurement and unitary evolution. In this paper, we investigate the relaxation critical dynamics near the MIPT for different initial states in a one-dimensional quantum circuit. Specifically, when the initial state is in the volume-law phase with vanishing measurement probability, we find that the half-chain entanglement entropy $S$ decays as $S\propto t^{-1}$ with the coefficients proportional to the size of the system in the short-time stage; In contrast, when the initial state is the product state, $S$ increases with time as $S\propto \ln{t}$, consistent with previous studies. Despite these contrasting behaviors, we develop a unified scaling form to describe these scaling behaviors for different initial states where the off-critical-point effects can also be incorporated. This framework offers significant advantages for experimental MIPT detection. Our novel scheme, leveraging relaxation dynamical scaling, drastically reduces post-selection overhead, and can eliminate it completely with trackable classical simulation.
\end{abstract}

\maketitle

{\it Introduction}--- Monitored many-body systems subjected to unitary evolution with interspersed projective measurements can exhibit dynamical phase transitions between the entangling phase and disentangling phase by tuning the probability of the measurements~\cite{LiydPhysRevB98205136,NahumPhysRevX9031009,LiydPhysRevB100134306}. This phenomenon is called the measurement-induced phase transition (MIPT)~\cite{LiydPhysRevB98205136,NahumPhysRevX9031009,LiydPhysRevB100134306}. Different from conventional phase transitions, which are usually characterized by the appearance of a finite order parameter, MIPT is distinguished by distinct properties of the entanglement entropy $S$ on two sides of the transition point. When the measurement probability is relatively low, unitary evolution dominates and renders the entanglement entropy $S$ conform to a volume-law scaling~\cite{Calabrese_2005,HusePhysRevLett111127205,NahumPhysRevX7031016,NahumPhysRevX8021014,SondhiPhysRevX8021013,TurkeshiPhysRevLett131230403}. On the other hand, for higher measurement probability, local measurements dominate and impede the spread of the information to local degrees of freedom by collapsing the wave function, such that the entanglement entropy obeys the area-law scaling ~\cite{Daviesbook,Zurekbook}. These unique features have propelled the MIPT into a prominent position, galvanizing a flurry of research activities in recent years. The continuous commitment to exploring the MIPT is anticipated to bear fruitful results, considering not only its profound theoretical significance, such as its intimate connections with percolation and conformal field theory, but also its practical relevance. Indeed, the MIPT is believed to be commonplace in the modern quantum devices, making it a subject of great interest and potential~\cite{LiydPhysRevB98205136,NahumPhysRevX9031009,LiydPhysRevB100134306,ChanPhysRevB99224307,VasseurPhysRevB100134203,HusePhysRevX10041020,HusePhysRevLett125070606,HusePhysRevB101060301,ChoiPhysRevLett125030505,JiancmPhysRevB101104302,BaoPhysRevB101104301,LiusPhysRevB107L201113,IppolitiPhysRevX11011030,TurkeshiPhysRevB106214316,TurkeshiPhysRevB102014315,LiydPhysRevB109174307,LiusPhysRevB110064323,LiusPhysRevLett132240402,CecilePhysRevResearch6033220,XuckPRXQuantum4030317,AshidaPhysRevB110094404,AliceaPhysRevX13041042,MoghaddamPhysRevLett131020401,XuckPhysRevB109035146,FisherPhysRevLett130220404,PolmannPRXQuantum5010309,Lavasaninpexp,Noelnpexp,Koh2023,Hoke2023,PhysRevLett.129.080501, PhysRevResearch.3.023200,kelly2024generalizingmeasurementinducedphasetransitions,Chen2024_z,qian2024coherentinformationphasetransition, PhysRevB.107.094309, yu2025gapless}.

In conventional phase transitions, universal critical phenomena appear not merely in long-time steady states~\cite{Hohenberg1977rmp} but also during short-time nonequilibrium relaxation processes~\cite{Janssen1989}. Owing to critical slowing down, the information encapsulated within the initial state can be retained over a macroscopic timescale. The fascinating interplay between this memory effect and the long-wavelength critical fluctuations gives rise to a rich tapestry of universal relaxation scaling behaviors. Notably, in classical phase transitions, it has been demonstrated that for completely ordered initial states, the evolution of the order parameter can be aptly characterized by a power function with its exponent being a combination of the equilibrium exponents and the dynamic exponent $z$~\cite{Janssen1989,Lizb1995prl,Zhengb1996prl,Zhengb1998review,Albano2011iop,Shuyr2024prb}. Analogous scaling properties have also emerged in quantum phase transitions~\cite{Polkovnikov2013prl,Schmalian2014prl,Schmalian2015prb,Chiocchetta2015prb,Marino201prl,Yin2019prl,Yin2021prb,Marino_2022,Yins2014prb,Yins2014pre,Shu2017prb,Yin2022prl,Yin2022prb,Zuo2021prb,Yin2024prbcor,Changwx2024}. It was shown that such dynamic scaling theory can be used to determine critical properties efficiently in various systems, even including sign-problematic fermion models~\cite{Yuarxiv2023,Yuarxiv2024}. Moreover, the relaxation critical dynamics in quantum phase transitions has been realized in experiment~\cite{Zhangsx2024prbys}.

To date, the majority of investigations concerning the MIPT have predominantly centered around the characteristics of steady states. Although there have been several pioneering studies~\cite{LiydPhysRevB98205136,NahumPhysRevX9031009}, research on non-equilibrium dynamics remains an area that calls for further exploration and advancement~\cite{Yinarxiv2024}. In particular, a crucial question emerges: for volume-law initial states, does the relaxation dynamics of entanglement entropy $S$ exhibit universal scaling behaviors? And furthermore, can scaling behaviors for different initial states be encapsulated by a unified scaling form?

To answer these questions, we delve into the relaxation critical dynamics of MIPT within a $(1+1)$D hybrid stabilizer circuit~\cite{LiydPhysRevB98205136,LiydPhysRevB100134306}. We prepare two distinct initial states: the volume-law state, which emerges as the steady state without measurement, and the product state, which can be regarded as the steady state with full measurement. Near the critical point, during the relaxation process, we uncover a plethora of scaling behaviors. For the volume-law initial state, we find a new dynamic scaling relation: the half-chain entanglement entropy $S$ decays as $S\propto t^{-1}$, with the coefficient proportional to the system size. In contrast, for the product initial state, we confirm that $S$ increases with time according to $S\propto \ln{t}$, consistent with previous reports~\cite{NahumPhysRevX9031009}.
Moreover, we find that these scaling behaviors can be described by a unified scaling form, in which the scaling function shows different asymptotic behaviors for different initial states. Additionally, we can incorporate off-critical-point effects into this scaling form.
We also point out that our findings can facilitate the experimental research of MIPT~\cite{Noelnpexp,Koh2023,Hoke2023}.

\begin{figure}[tbp]
\centering
  \includegraphics[width=\linewidth,clip]{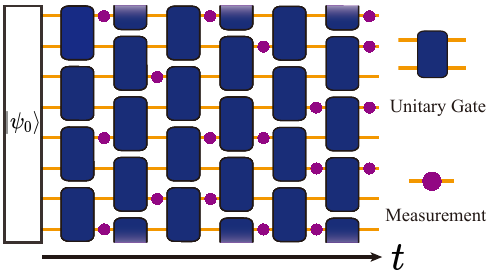}
  \vskip-3mm
  \caption{{\bf Sketch of relaxation critical dynamics in MIPT.} Dark blue rectangles represent two-site random Clifford gates and purple circles denote local projective measurements that are performed with a probability $p$. $|\psi(0)\rangle$ corresponds to the initial state, which is chosen as the steady state of random quantum circuits and the product state. The unit of time is set as two evolution steps.}
  \label{fig:quench}
\end{figure}


{\it Model and steady-state properties}--- We focus on the $(1+1)$D circuits with the regular brick-wall structure, as illustrated in Fig.~\ref{fig:quench}. The evolution can be decomposed into the unitary layers and the measurement layers. The unitary layer is made up of the random Clifford gates, which in each step are given by $U(t)=\left[\prod_{x-{\rm odd}} U_{(x,x+1),2t}\right]\left[\prod_{x-{\rm even}} U_{(x,x+1),2t+1}\right]$ starting from $t=0$. In addition, the measurement layer consists of the projective measurements on each qubit with probability $p$. The projector operator $P^{\pm}\equiv (1 \pm Z)/2$ is randomly applied to the spatial wave function according to the Born rules, followed by a normalization procedure~\cite{LiydPhysRevB98205136,LiydPhysRevB100134306}. The periodic boundary condition is imposed in the spatial direction.


As a paradigmatic model hosting the MIPT, the critical properties of the steady state of this model have been extensively explored. For large $p$, the projective measurement dominates and the steady state is in the area-law phase. For small $p$, the unitary evolution dominates, and the steady state is in the volume-law phase. In between, there is a critical point $p_c=0.159 95(10)$~\cite{TurkeshiPhysRevB106214316}. Near the MIPT, the half-chain entanglement entropy $S$ obeys a scaling form~\cite{LiydPhysRevB100134306} \begin{equation}
S(p,L)=\alpha \ln{L}+F(gL^{1/\nu}),
\label{steadyS}
\end{equation}
in which $g=p-p_c$, $\alpha=1.57(1)$~\cite{TurkeshiPhysRevB106214316}, $\nu=1.260(15)$~\cite{TurkeshiPhysRevB106214316}, and $F$ is the scaling function. This scaling form demonstrates that at the critical point $g=0$, $S\propto {\ln L}$. Moreover, in the area-law phase $S$ tends to a constant as $S\propto {\rm ln}\xi \propto {\rm ln}(g^{-\nu})$ when $g>L^{-1/\nu}$, whereas in the volume-law phase, $g<0$, $S\propto L$ and $F(gL^{1/\nu})$ satisfies $F(gL^{1/\nu})\propto (gL^{1/\nu})^{\nu}$~\cite{LiydPhysRevB98205136,LiydPhysRevB100134306}.

{\it General relaxation scaling form}--- Now we consider the relaxation critical dynamics of the MIPT for above circuit model, as illustrated in Fig.~\ref{fig:quench}. For the nonequilibrium process, the time $t$ with its exponent $z=1$ must be an indispensable scaling variable. Moreover, in contrast to the equilibrium case, which only depends on the measurement probability $p$, the relaxation dynamics also depends on the initial states as the result of the divergence of the correlation time. Accordingly, by incorporating the steady-state scaling form~(\ref{steadyS})~\cite{LiydPhysRevB100134306}, we obtain the general relaxation scaling form of $S$ as
\begin{equation}
S(t,g,L)=\alpha {\rm ln}L+G(tL^{-z},gL^{1/\nu},{X_0}),
\label{general}
\end{equation}
in which $X_0$ represents the initial information. For $t\rightarrow\infty$, Eq.~(\ref{general1}) recovers the scaling form of the steady state, i.e., Eq.~(\ref{general}). Although here the half-chain entanglement entropy is mainly studied, Eq.~(\ref{general}) can readily be generalized to the case for the entanglement entropy defined on a subsystem with size $A$, wherein an additional scaling variable $AL^{-1}$ should be included~\cite{Yinarxiv2024}.

\begin{figure}[tbp]
\centering
  \includegraphics[width=\linewidth,clip]{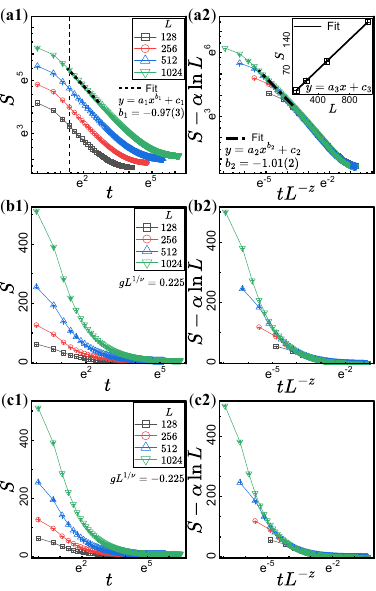}
  \vskip-3mm
  \caption{{\bf Relaxation dynamics of entanglement entropy $S$ from the volume-law initial state.} (a) At $g=0$, curves of $S$ versus $t$ before (a1) and after (a2) rescaling for different sizes. The inset in (a2) shows the linear dependence of $S$ versus $L$ for the time marked (vertical dashed line) in (a1). Double logarithmic scales are used in (a). (b) Curves of $S$ versus $t$ with a fixed $gL^{1/\nu}>0$ for different $L$ before (b1) and after (b2) rescaling. (c) Curves of $S$ versus $t$ with a fixed $gL^{1/\nu}<0$ for different $L$ before (c1) and after (c2) rescaling. Semi-logarithmic scales are used in (b) and (c).
  }
  \label{fig:quench1}
\end{figure}

In this paper, we focus on two kinds of initial states: one is the volume-law state prepared as the steady state for $p=0$, and the other is the product state corresponds to the steady state for $p=1$. Note that both $p=0$ and $p=1$ are the fixed points of the scale transformation of $p$. Accordingly, for these states, Eq.~(\ref{general}) can be simplified as
\begin{equation}
S(t,g,L)=\alpha {\ln}L+G_1(tL^{-z},gL^{1/\nu}),
\label{general1}
\end{equation}
in which $X_0$ disappears and $G_1$ depends on the initial states implicitly.

{\it Relaxation dynamics from volume-law initial state}---We start with the relaxation dynamics from the volume-law steady state for $p=0$. We first focus on the case of $g=0$. In this case, Eq.~(\ref{general1}) reduces to
\begin{equation}
S(t,L)=\alpha {\rm ln}L+G_2(tL^{-z}).
\label{volume1}
\end{equation}
Fig.~\ref{fig:quench1} (a1) shows the numerical results of the evolution of $S$ for different $L$. In Fig.~\ref{fig:quench1} (a2), we plot the curves of $(S-\alpha{\ln}L)$ versus $tL^{-z}$. We find that these curves collapse well onto a single curve, confirming that the relaxation dynamics of $S$ can be described by Eq.~(\ref{volume1}).

Next, we explore the short-time dynamics of $S$ for this initial state. As illustrated in Fig.~\ref{fig:quench1}(a1), in the short-time stage, the evolution of $S$ conforms to a power function, and the exponents are fairly similar for all $L$. Fitting the curve for $L=1024$ demonstrates that the exponent is close to one, which in turn leads to the short-time scaling relation for $S$,
\begin{equation}
S(t,L)\propto t^{-1}.
\label{volume2}
\end{equation}

\begin{figure}[tbp]
\centering
  \includegraphics[width=\linewidth,clip]{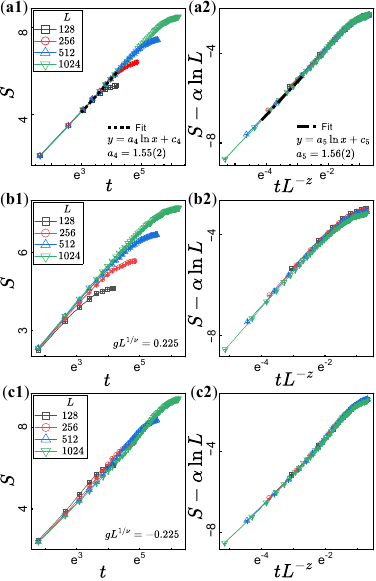}
  \vskip-3mm
  \caption{{\bf Relaxation dynamics of entanglement entropy $S$ from the product initial state.} (a) At $g=0$, curves of $S$ versus $t$ before (a1) and after (a2) rescaling for different sizes. (b) Curves of $S$ versus $t$ with a fixed $gL^{1/\nu}>0$ for different $L$ before (b1) and after (b2) rescaling. (c) Curves of $S$ versus $t$ with a fixed $gL^{1/\nu}<0$ for different $L$ before (c1) and after (c2) rescaling. Semi-logarithmic scales are used in all figures.
  }
  \label{fig:quench2}
\end{figure}

This scaling relation, Eq.~(\ref{volume2}), requires that $G_2$ for small $t$ must obey
\begin{equation}
G_2(tL^{-z})\sim (tL^{-z})^{-1/z}\sim Lt^{-1},
\label{volume3}
\end{equation}
in which $z=1$ has been set as input. In Fig.~\ref{fig:quench2} (b2), by fitting the rescaled curve, which is just the curve of $G_2(tL^{-z})$, with a power function in the short-time stage, we find that the exponent is quite close to one. Moreover, for a fixed $t$ in the short-time stage, as marked in Fig.~\ref{fig:quench1} (a1), we find that $S\propto L$ as shown in the inset of Fig.~\ref{fig:quench1} (a2). These results confirm the asymptotic form of $G_2(tL^{-z})$.

To understand the physical reason for Eqs.~(\ref{volume2}) and (\ref{volume3}), we note that $S$ in the initial state satisfies the volume law. At the critical point, as a result of critical slowing down, this volume-law characteristic endures during the macroscopic short-time stage. This dictates that $G_2(tL^{-z})$ must satisfy $G_2(tL^{-z})\propto L$ and dominates over the logarithmic term, thus leading to Eq.~(\ref{volume2}). Accordingly, Eqs.~(\ref{volume2}) and (\ref{volume3}) vividly reflect the consequence of the interplay between the volume-law initial state and the divergent relaxation time at the critical point.

Then we turn our attention to the examination of Eq.~(\ref{general1})  for $g\neq0$. As shown in Figs.~\ref{fig:quench1} (b) and (c), for a fixed $gL^{1/\nu}$ [noting that $g>0$ in (b) and $g<0$ in (c)],  it can be observed that the rescaled curves of $(S-\alpha{\ln}L)$ versus $tL^{-z}$ collapse onto each other well. These results firmly verify that Eq.~(\ref{general1}) provides an apt description on the relation critical dynamics of $S$ from the steady state for $p=0$.

{\it Relaxation dynamics from product initial state}---Now we turn to the relaxation critical dynamics of $S$ from a product initial state. At first, we study the relaxation scaling property of $S$ at $p_c$. In this case, Eq.~(\ref{general1}) reduces to
\begin{equation}
S(t,L)=\alpha {\rm ln}L+G_3(tL^{-z}).
\label{area1}
\end{equation}
Although it seems that Eq.~(\ref{area1}) is similar to Eq.~(\ref{volume1}), we will find that the scaling functions are remarkably different for two cases. Fig.~\ref{fig:quench2} (a1) shows the evolution of $S$ for different $L$. By rescaling the $S$ and $t$ as $(S-\alpha{\ln}L)$ versus $tL^{-z}$, respectively, we find that the rescaled curves match well with each other, confirming Eq.~(\ref{area1}).

Then we investigate the short-time dynamics of $S$. Fig.~\ref{fig:quench2} (a1) shows that in the short-time stage, the growth of $S$ versus $t$ obeys $S\propto \ln t$ (straight line in semi-logarithmic scale) and scarcely depends on $L$. This requires $G_3$ to obey
\begin{equation}
G_3(tL^{-z})= \delta {\rm ln}(tL^{-z})+c,
\label{area2}
\end{equation}
in which $c$ is a nonuniversal constant and $\delta=\alpha/z$ such that the $L$-dependent term can be cancelled. Accordingly, one can obtain the scaling relation of $S$ in the short-time stage,
\begin{equation}
S(t,L)= \delta {\rm ln}t+c.
\label{area3}
\end{equation}
As shown in Fig.~\ref{fig:quench2} (a1), by fitting the curve of $S$ versus $t$ for $L=1024$ with a logarithmic function, we obtain $\delta=1.55(2)$, which is consistent with the value of $\alpha$ within the error bar, confirming Eq.~(\ref{area3}). In addition, the curve of $(S-\alpha{\ln}L)$ versus $tL^{-z}$ displayed in Fig.~\ref{fig:quench2} (a2) just corresponds to the scaling function $G_3(tL^{-z})$. The logarithmic fitting on this curve demonstrates that the coefficient of ${\rm ln}(tL^{-z})$ is also close to the value of $\alpha$, confirming Eq.~(\ref{area2}).

Note that as an archetypal setup of the MIPT, the dynamics of $S$ from the product state was also studied in previous works~\cite{LiydPhysRevB98205136,NahumPhysRevX9031009} and the logarithmic growth of $S$ was also discovered~\cite{NahumPhysRevX9031009}. Here, the discussion we presented above give a more systematic understanding based on the dynamic scaling analyses.

For the off-critical-point cases with $g\neq0$, we examine Eq.~(\ref{general1}) in Fig.~\ref{fig:quench2} (b) and (c). For a fixed $gL^{1/\nu}$ [$g>0$ in Fig.~\ref{fig:quench2}(b) and $g<0$ in Fig.~\ref{fig:quench2}(c)], we find that the rescaled curves of $(S-\alpha{\ln}L)$ versus $tL^{-z}$ collapse well onto each other, confirming that Eq.~(\ref{general1}) provides a unified description for both the volume-law initial state and the product initial state.

Moreover, Fig.~\ref{fig:quench2} demonstrates that for all $g$ close to the critical point, the evolution of $S$ in the short-time stage can be approximated as $S(t,L)\propto {\rm ln}t$ and the finite-size effect is weak. The reason for this lies in the fact that, starting from a product initial state, the correlation length $\xi$ increases as $\xi\sim t^{1/z}$ in the short-time stage. When $\xi$ is much smaller than $L$, the size effects can be disregarded. Consequently, we can obtain the short-time scaling form of $S$ from a product initial state as
\begin{equation}
S(t)= \delta {\rm ln}t+G_4(gt^{1/\nu z}).
\label{area4}
\end{equation}

\begin{figure}[tbp]
\centering
  \includegraphics[width=\linewidth,clip]{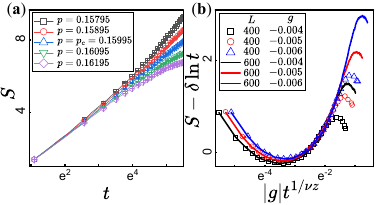}
  \vskip-3mm
  \caption{{\bf Short-time dynamics of entanglement entropy $S$ from product initial state.} (a) Curves of $S$ versus $t$ for different $p$ near the critical point. (b) Curves of $S-\delta \ln t$ versus $gt^{1/\nu z}$ for $L=400$ and $L=600$. Semi-logarithmic scales are used in both figures.
  }
  \label{fig:quench3}
\end{figure}

Eq.~(\ref{area4}) offers an effective way to determine the critical properties of MIPT. To illustrate, we compare the curves of $S$ versus $t$ in semi-logarithmic scale for different $p$ near the critical point in Fig.~\ref{fig:quench3} (a). We observe that at the critical point, the curve is a straight line, while on different sides of MIPT, the curves show distinct trends: downward for $p>p_c$ and upward for $p<p_c$. Thus, using this approach, we can identify the critical point in the short-time stage with a much lower cost.

Moreover, for $p<p_c$, although it was shown that $S$ increases with $t$ as $S\propto t$ in the long-time stage and eventually saturates to $S\propto L$~\cite{NahumPhysRevX9031009}, we find that Eq.~(\ref{area4}) is still applicable in the short-time critical region. Fig.~\ref{fig:quench3} (b) confirms this conclusion by showing that for different sizes the rescaled curves of $S$ versus $t$ collapse well according to Eq.~(\ref{area4}) after a transient time scale in the short-time stage for $g<0$, although apparent deviations appear in the long-time stage.

{\it Discussion}---In experimental research on MIPTs, the post-selection problem is a major hurdle as the probability of identical trajectories decays exponentially with measurement number~\cite{Noelnpexp,Koh2023,Hoke2023}. In other words, the overhead associated with the post-selection complexity is proportional to $\exp{(pLt)}$. This means that to track the steady-state properties, the number of independent experiments has to reach the magnitude level of $\exp{(pLt_{eq})}$ with $t_{eq}$ being the equilibration time. Given that we can explore the MIPT in the short-time scale, which is much less than $t_{eq}$, it is evident that the overhead of the post-selection can be reduced considerably.

Furthermore, our method can be combined with the cross-correlation protocol~\cite{AltmanPRXQuantum5030311} to completely eliminate the post-selection problem. The cross-correlation protocol in MIPT was devised as an ingenious method by estimating the entanglement entropy with appropriate cross-correlations which can be obtained with the help of classical simulations to mitigate the post-selection problem~\cite{AltmanPRXQuantum5030311}. However, the capacity of the classical computer still hinders the investigation of the volume-law regime since $S$ is proportional to $L$ in the steady state. Now, we have shown that Eq.~\eqref{area4} is applicable to $p<p_{c}$ in the short-time stage with much smaller $S$. Consequently, it is expected that the relaxation dynamic scaling, combined with the cross-correlation protocol, can facilitate the experimental study of the MIPT by alleviating the post-selection problem.


{\it Summary}--- In this paper, we have conducted an in-depth study on the relaxation critical dynamics of the MIPT in a $(1+1)$D hybrid stabilizer circuits. For relaxation dynamics from the volume-law initial state, our investigations have led to the discovery that the entanglement entropy $S$ satisfies $S\propto t^{-1}$ at $p_c$. In contrast, for relaxation dynamics from the product initial state, we have verified that $S$ at $p_c$ satisfies $S\propto \ln t$, which aligns well with the findings of previous works. Despite these differences, we have confirmed that the dynamics of the entanglement entropy $S$ can be aptly characterized by a unified relaxation scaling form, in which the scaling function has different asymptotic properties for different initial states. Moreover, we have discussed that the present work offers a feasible approach to investigate the MIPT on realistic quantum devices by significantly mitigating the post-selection problem~~\cite{Noelnpexp,Koh2023,Hoke2023}.

{\it Acknowledgments}---W.W., J.L., and S.Y. are supported by the National Natural Science Foundation of China (Grants No. 12222515 and No. 12075324). S.Y. is also supported by the Science and Technology Projects in Guangdong Province (Grants No. 2021QN02X561). S.-X. Z. is supported by a startup grant at IOP-CAS.

\bibliographystyle{apsrev4-1}
\let\oldaddcontentsline\addcontentsline
\renewcommand{\addcontentsline}[3]{}
\bibliography{ref.bib}

\begin{thebibliography}{73}%
\makeatletter
\providecommand \@ifxundefined [1]{%
 \@ifx{#1\undefined}
}%
\providecommand \@ifnum [1]{%
 \ifnum #1\expandafter \@firstoftwo
 \else \expandafter \@secondoftwo
 \fi
}%
\providecommand \@ifx [1]{%
 \ifx #1\expandafter \@firstoftwo
 \else \expandafter \@secondoftwo
 \fi
}%
\providecommand \natexlab [1]{#1}%
\providecommand \enquote  [1]{``#1''}%
\providecommand \bibnamefont  [1]{#1}%
\providecommand \bibfnamefont [1]{#1}%
\providecommand \citenamefont [1]{#1}%
\providecommand \href@noop [0]{\@secondoftwo}%
\providecommand \href [0]{\begingroup \@sanitize@url \@href}%
\providecommand \@href[1]{\@@startlink{#1}\@@href}%
\providecommand \@@href[1]{\endgroup#1\@@endlink}%
\providecommand \@sanitize@url [0]{\catcode `\\12\catcode `\$12\catcode `\&12\catcode `\#12\catcode `\^12\catcode `\_12\catcode `\%12\relax}%
\providecommand \@@startlink[1]{}%
\providecommand \@@endlink[0]{}%
\providecommand \url  [0]{\begingroup\@sanitize@url \@url }%
\providecommand \@url [1]{\endgroup\@href {#1}{\urlprefix }}%
\providecommand \urlprefix  [0]{URL }%
\providecommand \Eprint [0]{\href }%
\providecommand \doibase [0]{http://dx.doi.org/}%
\providecommand \selectlanguage [0]{\@gobble}%
\providecommand \bibinfo  [0]{\@secondoftwo}%
\providecommand \bibfield  [0]{\@secondoftwo}%
\providecommand \translation [1]{[#1]}%
\providecommand \BibitemOpen [0]{}%
\providecommand \bibitemStop [0]{}%
\providecommand \bibitemNoStop [0]{.\EOS\space}%
\providecommand \EOS [0]{\spacefactor3000\relax}%
\providecommand \BibitemShut  [1]{\csname bibitem#1\endcsname}%
\let\auto@bib@innerbib\@empty
\bibitem [{\citenamefont {Li}\ \emph {et~al.}(2018)\citenamefont {Li}, \citenamefont {Chen},\ and\ \citenamefont {Fisher}}]{LiydPhysRevB98205136}%
  \BibitemOpen
  \bibfield  {author} {\bibinfo {author} {\bibfnamefont {Y.}~\bibnamefont {Li}}, \bibinfo {author} {\bibfnamefont {X.}~\bibnamefont {Chen}}, \ and\ \bibinfo {author} {\bibfnamefont {M.~P.~A.}\ \bibnamefont {Fisher}},\ }\href {\doibase 10.1103/PhysRevB.98.205136} {\bibfield  {journal} {\bibinfo  {journal} {Phys. Rev. B}\ }\textbf {\bibinfo {volume} {98}},\ \bibinfo {pages} {205136} (\bibinfo {year} {2018})}\BibitemShut {NoStop}%
\bibitem [{\citenamefont {Skinner}\ \emph {et~al.}(2019)\citenamefont {Skinner}, \citenamefont {Ruhman},\ and\ \citenamefont {Nahum}}]{NahumPhysRevX9031009}%
  \BibitemOpen
  \bibfield  {author} {\bibinfo {author} {\bibfnamefont {B.}~\bibnamefont {Skinner}}, \bibinfo {author} {\bibfnamefont {J.}~\bibnamefont {Ruhman}}, \ and\ \bibinfo {author} {\bibfnamefont {A.}~\bibnamefont {Nahum}},\ }\href {\doibase 10.1103/PhysRevX.9.031009} {\bibfield  {journal} {\bibinfo  {journal} {Phys. Rev. X}\ }\textbf {\bibinfo {volume} {9}},\ \bibinfo {pages} {031009} (\bibinfo {year} {2019})}\BibitemShut {NoStop}%
\bibitem [{\citenamefont {Li}\ \emph {et~al.}(2019)\citenamefont {Li}, \citenamefont {Chen},\ and\ \citenamefont {Fisher}}]{LiydPhysRevB100134306}%
  \BibitemOpen
  \bibfield  {author} {\bibinfo {author} {\bibfnamefont {Y.}~\bibnamefont {Li}}, \bibinfo {author} {\bibfnamefont {X.}~\bibnamefont {Chen}}, \ and\ \bibinfo {author} {\bibfnamefont {M.~P.~A.}\ \bibnamefont {Fisher}},\ }\href {\doibase 10.1103/PhysRevB.100.134306} {\bibfield  {journal} {\bibinfo  {journal} {Phys. Rev. B}\ }\textbf {\bibinfo {volume} {100}},\ \bibinfo {pages} {134306} (\bibinfo {year} {2019})}\BibitemShut {NoStop}%
\bibitem [{\citenamefont {Calabrese}\ and\ \citenamefont {Cardy}(2005)}]{Calabrese_2005}%
  \BibitemOpen
  \bibfield  {author} {\bibinfo {author} {\bibfnamefont {P.}~\bibnamefont {Calabrese}}\ and\ \bibinfo {author} {\bibfnamefont {J.}~\bibnamefont {Cardy}},\ }\href {\doibase 10.1088/1742-5468/2005/04/P04010} {\bibfield  {journal} {\bibinfo  {journal} {Journal of Statistical Mechanics: Theory and Experiment}\ }\textbf {\bibinfo {volume} {2005}},\ \bibinfo {pages} {P04010} (\bibinfo {year} {2005})}\BibitemShut {NoStop}%
\bibitem [{\citenamefont {Kim}\ and\ \citenamefont {Huse}(2013)}]{HusePhysRevLett111127205}%
  \BibitemOpen
  \bibfield  {author} {\bibinfo {author} {\bibfnamefont {H.}~\bibnamefont {Kim}}\ and\ \bibinfo {author} {\bibfnamefont {D.~A.}\ \bibnamefont {Huse}},\ }\href {\doibase 10.1103/PhysRevLett.111.127205} {\bibfield  {journal} {\bibinfo  {journal} {Phys. Rev. Lett.}\ }\textbf {\bibinfo {volume} {111}},\ \bibinfo {pages} {127205} (\bibinfo {year} {2013})}\BibitemShut {NoStop}%
\bibitem [{\citenamefont {Nahum}\ \emph {et~al.}(2017)\citenamefont {Nahum}, \citenamefont {Ruhman}, \citenamefont {Vijay},\ and\ \citenamefont {Haah}}]{NahumPhysRevX7031016}%
  \BibitemOpen
  \bibfield  {author} {\bibinfo {author} {\bibfnamefont {A.}~\bibnamefont {Nahum}}, \bibinfo {author} {\bibfnamefont {J.}~\bibnamefont {Ruhman}}, \bibinfo {author} {\bibfnamefont {S.}~\bibnamefont {Vijay}}, \ and\ \bibinfo {author} {\bibfnamefont {J.}~\bibnamefont {Haah}},\ }\href {\doibase 10.1103/PhysRevX.7.031016} {\bibfield  {journal} {\bibinfo  {journal} {Phys. Rev. X}\ }\textbf {\bibinfo {volume} {7}},\ \bibinfo {pages} {031016} (\bibinfo {year} {2017})}\BibitemShut {NoStop}%
\bibitem [{\citenamefont {Nahum}\ \emph {et~al.}(2018)\citenamefont {Nahum}, \citenamefont {Vijay},\ and\ \citenamefont {Haah}}]{NahumPhysRevX8021014}%
  \BibitemOpen
  \bibfield  {author} {\bibinfo {author} {\bibfnamefont {A.}~\bibnamefont {Nahum}}, \bibinfo {author} {\bibfnamefont {S.}~\bibnamefont {Vijay}}, \ and\ \bibinfo {author} {\bibfnamefont {J.}~\bibnamefont {Haah}},\ }\href {\doibase 10.1103/PhysRevX.8.021014} {\bibfield  {journal} {\bibinfo  {journal} {Phys. Rev. X}\ }\textbf {\bibinfo {volume} {8}},\ \bibinfo {pages} {021014} (\bibinfo {year} {2018})}\BibitemShut {NoStop}%
\bibitem [{\citenamefont {von Keyserlingk}\ \emph {et~al.}(2018)\citenamefont {von Keyserlingk}, \citenamefont {Rakovszky}, \citenamefont {Pollmann},\ and\ \citenamefont {Sondhi}}]{SondhiPhysRevX8021013}%
  \BibitemOpen
  \bibfield  {author} {\bibinfo {author} {\bibfnamefont {C.~W.}\ \bibnamefont {von Keyserlingk}}, \bibinfo {author} {\bibfnamefont {T.}~\bibnamefont {Rakovszky}}, \bibinfo {author} {\bibfnamefont {F.}~\bibnamefont {Pollmann}}, \ and\ \bibinfo {author} {\bibfnamefont {S.~L.}\ \bibnamefont {Sondhi}},\ }\href {\doibase 10.1103/PhysRevX.8.021013} {\bibfield  {journal} {\bibinfo  {journal} {Phys. Rev. X}\ }\textbf {\bibinfo {volume} {8}},\ \bibinfo {pages} {021013} (\bibinfo {year} {2018})}\BibitemShut {NoStop}%
\bibitem [{\citenamefont {Sierant}\ \emph {et~al.}(2023)\citenamefont {Sierant}, \citenamefont {Schir\`o}, \citenamefont {Lewenstein},\ and\ \citenamefont {Turkeshi}}]{TurkeshiPhysRevLett131230403}%
  \BibitemOpen
  \bibfield  {author} {\bibinfo {author} {\bibfnamefont {P.}~\bibnamefont {Sierant}}, \bibinfo {author} {\bibfnamefont {M.}~\bibnamefont {Schir\`o}}, \bibinfo {author} {\bibfnamefont {M.}~\bibnamefont {Lewenstein}}, \ and\ \bibinfo {author} {\bibfnamefont {X.}~\bibnamefont {Turkeshi}},\ }\href {\doibase 10.1103/PhysRevLett.131.230403} {\bibfield  {journal} {\bibinfo  {journal} {Phys. Rev. Lett.}\ }\textbf {\bibinfo {volume} {131}},\ \bibinfo {pages} {230403} (\bibinfo {year} {2023})}\BibitemShut {NoStop}%
\bibitem [{\citenamefont {Davies}(1976)}]{Daviesbook}%
  \BibitemOpen
  \bibfield  {author} {\bibinfo {author} {\bibfnamefont {E.~B.}\ \bibnamefont {Davies}},\ }\href@noop {} {\emph {\bibinfo {title} {Quantum theory of open systems}}}\ (\bibinfo  {publisher} {Academic Press London},\ \bibinfo {year} {1976})\BibitemShut {NoStop}%
\bibitem [{\citenamefont {Wheeler}\ and\ \citenamefont {Zurek}(2014)}]{Zurekbook}%
  \BibitemOpen
  \bibfield  {author} {\bibinfo {author} {\bibfnamefont {J.~A.}\ \bibnamefont {Wheeler}}\ and\ \bibinfo {author} {\bibfnamefont {W.~H.}\ \bibnamefont {Zurek}},\ }\href@noop {} {\emph {\bibinfo {title} {Quantum theory and measurement}}},\ Vol.~\bibinfo {volume} {81}\ (\bibinfo  {publisher} {Princeton University Press},\ \bibinfo {year} {2014})\BibitemShut {NoStop}%
\bibitem [{\citenamefont {Chan}\ \emph {et~al.}(2019)\citenamefont {Chan}, \citenamefont {Nandkishore}, \citenamefont {Pretko},\ and\ \citenamefont {Smith}}]{ChanPhysRevB99224307}%
  \BibitemOpen
  \bibfield  {author} {\bibinfo {author} {\bibfnamefont {A.}~\bibnamefont {Chan}}, \bibinfo {author} {\bibfnamefont {R.~M.}\ \bibnamefont {Nandkishore}}, \bibinfo {author} {\bibfnamefont {M.}~\bibnamefont {Pretko}}, \ and\ \bibinfo {author} {\bibfnamefont {G.}~\bibnamefont {Smith}},\ }\href {\doibase 10.1103/PhysRevB.99.224307} {\bibfield  {journal} {\bibinfo  {journal} {Phys. Rev. B}\ }\textbf {\bibinfo {volume} {99}},\ \bibinfo {pages} {224307} (\bibinfo {year} {2019})}\BibitemShut {NoStop}%
\bibitem [{\citenamefont {Vasseur}\ \emph {et~al.}(2019)\citenamefont {Vasseur}, \citenamefont {Potter}, \citenamefont {You},\ and\ \citenamefont {Ludwig}}]{VasseurPhysRevB100134203}%
  \BibitemOpen
  \bibfield  {author} {\bibinfo {author} {\bibfnamefont {R.}~\bibnamefont {Vasseur}}, \bibinfo {author} {\bibfnamefont {A.~C.}\ \bibnamefont {Potter}}, \bibinfo {author} {\bibfnamefont {Y.-Z.}\ \bibnamefont {You}}, \ and\ \bibinfo {author} {\bibfnamefont {A.~W.~W.}\ \bibnamefont {Ludwig}},\ }\href {\doibase 10.1103/PhysRevB.100.134203} {\bibfield  {journal} {\bibinfo  {journal} {Phys. Rev. B}\ }\textbf {\bibinfo {volume} {100}},\ \bibinfo {pages} {134203} (\bibinfo {year} {2019})}\BibitemShut {NoStop}%
\bibitem [{\citenamefont {Gullans}\ and\ \citenamefont {Huse}(2020{\natexlab{a}})}]{HusePhysRevX10041020}%
  \BibitemOpen
  \bibfield  {author} {\bibinfo {author} {\bibfnamefont {M.~J.}\ \bibnamefont {Gullans}}\ and\ \bibinfo {author} {\bibfnamefont {D.~A.}\ \bibnamefont {Huse}},\ }\href {\doibase 10.1103/PhysRevX.10.041020} {\bibfield  {journal} {\bibinfo  {journal} {Phys. Rev. X}\ }\textbf {\bibinfo {volume} {10}},\ \bibinfo {pages} {041020} (\bibinfo {year} {2020}{\natexlab{a}})}\BibitemShut {NoStop}%
\bibitem [{\citenamefont {Gullans}\ and\ \citenamefont {Huse}(2020{\natexlab{b}})}]{HusePhysRevLett125070606}%
  \BibitemOpen
  \bibfield  {author} {\bibinfo {author} {\bibfnamefont {M.~J.}\ \bibnamefont {Gullans}}\ and\ \bibinfo {author} {\bibfnamefont {D.~A.}\ \bibnamefont {Huse}},\ }\href {\doibase 10.1103/PhysRevLett.125.070606} {\bibfield  {journal} {\bibinfo  {journal} {Phys. Rev. Lett.}\ }\textbf {\bibinfo {volume} {125}},\ \bibinfo {pages} {070606} (\bibinfo {year} {2020}{\natexlab{b}})}\BibitemShut {NoStop}%
\bibitem [{\citenamefont {Zabalo}\ \emph {et~al.}(2020)\citenamefont {Zabalo}, \citenamefont {Gullans}, \citenamefont {Wilson}, \citenamefont {Gopalakrishnan}, \citenamefont {Huse},\ and\ \citenamefont {Pixley}}]{HusePhysRevB101060301}%
  \BibitemOpen
  \bibfield  {author} {\bibinfo {author} {\bibfnamefont {A.}~\bibnamefont {Zabalo}}, \bibinfo {author} {\bibfnamefont {M.~J.}\ \bibnamefont {Gullans}}, \bibinfo {author} {\bibfnamefont {J.~H.}\ \bibnamefont {Wilson}}, \bibinfo {author} {\bibfnamefont {S.}~\bibnamefont {Gopalakrishnan}}, \bibinfo {author} {\bibfnamefont {D.~A.}\ \bibnamefont {Huse}}, \ and\ \bibinfo {author} {\bibfnamefont {J.~H.}\ \bibnamefont {Pixley}},\ }\href {\doibase 10.1103/PhysRevB.101.060301} {\bibfield  {journal} {\bibinfo  {journal} {Phys. Rev. B}\ }\textbf {\bibinfo {volume} {101}},\ \bibinfo {pages} {060301} (\bibinfo {year} {2020})}\BibitemShut {NoStop}%
\bibitem [{\citenamefont {Choi}\ \emph {et~al.}(2020)\citenamefont {Choi}, \citenamefont {Bao}, \citenamefont {Qi},\ and\ \citenamefont {Altman}}]{ChoiPhysRevLett125030505}%
  \BibitemOpen
  \bibfield  {author} {\bibinfo {author} {\bibfnamefont {S.}~\bibnamefont {Choi}}, \bibinfo {author} {\bibfnamefont {Y.}~\bibnamefont {Bao}}, \bibinfo {author} {\bibfnamefont {X.-L.}\ \bibnamefont {Qi}}, \ and\ \bibinfo {author} {\bibfnamefont {E.}~\bibnamefont {Altman}},\ }\href {\doibase 10.1103/PhysRevLett.125.030505} {\bibfield  {journal} {\bibinfo  {journal} {Phys. Rev. Lett.}\ }\textbf {\bibinfo {volume} {125}},\ \bibinfo {pages} {030505} (\bibinfo {year} {2020})}\BibitemShut {NoStop}%
\bibitem [{\citenamefont {Jian}\ \emph {et~al.}(2020)\citenamefont {Jian}, \citenamefont {You}, \citenamefont {Vasseur},\ and\ \citenamefont {Ludwig}}]{JiancmPhysRevB101104302}%
  \BibitemOpen
  \bibfield  {author} {\bibinfo {author} {\bibfnamefont {C.-M.}\ \bibnamefont {Jian}}, \bibinfo {author} {\bibfnamefont {Y.-Z.}\ \bibnamefont {You}}, \bibinfo {author} {\bibfnamefont {R.}~\bibnamefont {Vasseur}}, \ and\ \bibinfo {author} {\bibfnamefont {A.~W.~W.}\ \bibnamefont {Ludwig}},\ }\href {\doibase 10.1103/PhysRevB.101.104302} {\bibfield  {journal} {\bibinfo  {journal} {Phys. Rev. B}\ }\textbf {\bibinfo {volume} {101}},\ \bibinfo {pages} {104302} (\bibinfo {year} {2020})}\BibitemShut {NoStop}%
\bibitem [{\citenamefont {Bao}\ \emph {et~al.}(2020)\citenamefont {Bao}, \citenamefont {Choi},\ and\ \citenamefont {Altman}}]{BaoPhysRevB101104301}%
  \BibitemOpen
  \bibfield  {author} {\bibinfo {author} {\bibfnamefont {Y.}~\bibnamefont {Bao}}, \bibinfo {author} {\bibfnamefont {S.}~\bibnamefont {Choi}}, \ and\ \bibinfo {author} {\bibfnamefont {E.}~\bibnamefont {Altman}},\ }\href {\doibase 10.1103/PhysRevB.101.104301} {\bibfield  {journal} {\bibinfo  {journal} {Phys. Rev. B}\ }\textbf {\bibinfo {volume} {101}},\ \bibinfo {pages} {104301} (\bibinfo {year} {2020})}\BibitemShut {NoStop}%
\bibitem [{\citenamefont {Liu}\ \emph {et~al.}(2023)\citenamefont {Liu}, \citenamefont {Li}, \citenamefont {Zhang}, \citenamefont {Jian},\ and\ \citenamefont {Yao}}]{LiusPhysRevB107L201113}%
  \BibitemOpen
  \bibfield  {author} {\bibinfo {author} {\bibfnamefont {S.}~\bibnamefont {Liu}}, \bibinfo {author} {\bibfnamefont {M.-R.}\ \bibnamefont {Li}}, \bibinfo {author} {\bibfnamefont {S.-X.}\ \bibnamefont {Zhang}}, \bibinfo {author} {\bibfnamefont {S.-K.}\ \bibnamefont {Jian}}, \ and\ \bibinfo {author} {\bibfnamefont {H.}~\bibnamefont {Yao}},\ }\href {\doibase 10.1103/PhysRevB.107.L201113} {\bibfield  {journal} {\bibinfo  {journal} {Phys. Rev. B}\ }\textbf {\bibinfo {volume} {107}},\ \bibinfo {pages} {L201113} (\bibinfo {year} {2023})}\BibitemShut {NoStop}%
\bibitem [{\citenamefont {Ippoliti}\ \emph {et~al.}(2021)\citenamefont {Ippoliti}, \citenamefont {Gullans}, \citenamefont {Gopalakrishnan}, \citenamefont {Huse},\ and\ \citenamefont {Khemani}}]{IppolitiPhysRevX11011030}%
  \BibitemOpen
  \bibfield  {author} {\bibinfo {author} {\bibfnamefont {M.}~\bibnamefont {Ippoliti}}, \bibinfo {author} {\bibfnamefont {M.~J.}\ \bibnamefont {Gullans}}, \bibinfo {author} {\bibfnamefont {S.}~\bibnamefont {Gopalakrishnan}}, \bibinfo {author} {\bibfnamefont {D.~A.}\ \bibnamefont {Huse}}, \ and\ \bibinfo {author} {\bibfnamefont {V.}~\bibnamefont {Khemani}},\ }\href {\doibase 10.1103/PhysRevX.11.011030} {\bibfield  {journal} {\bibinfo  {journal} {Phys. Rev. X}\ }\textbf {\bibinfo {volume} {11}},\ \bibinfo {pages} {011030} (\bibinfo {year} {2021})}\BibitemShut {NoStop}%
\bibitem [{\citenamefont {Sierant}\ \emph {et~al.}(2022)\citenamefont {Sierant}, \citenamefont {Schir\`o}, \citenamefont {Lewenstein},\ and\ \citenamefont {Turkeshi}}]{TurkeshiPhysRevB106214316}%
  \BibitemOpen
  \bibfield  {author} {\bibinfo {author} {\bibfnamefont {P.}~\bibnamefont {Sierant}}, \bibinfo {author} {\bibfnamefont {M.}~\bibnamefont {Schir\`o}}, \bibinfo {author} {\bibfnamefont {M.}~\bibnamefont {Lewenstein}}, \ and\ \bibinfo {author} {\bibfnamefont {X.}~\bibnamefont {Turkeshi}},\ }\href {\doibase 10.1103/PhysRevB.106.214316} {\bibfield  {journal} {\bibinfo  {journal} {Phys. Rev. B}\ }\textbf {\bibinfo {volume} {106}},\ \bibinfo {pages} {214316} (\bibinfo {year} {2022})}\BibitemShut {NoStop}%
\bibitem [{\citenamefont {Turkeshi}\ \emph {et~al.}(2020)\citenamefont {Turkeshi}, \citenamefont {Fazio},\ and\ \citenamefont {Dalmonte}}]{TurkeshiPhysRevB102014315}%
  \BibitemOpen
  \bibfield  {author} {\bibinfo {author} {\bibfnamefont {X.}~\bibnamefont {Turkeshi}}, \bibinfo {author} {\bibfnamefont {R.}~\bibnamefont {Fazio}}, \ and\ \bibinfo {author} {\bibfnamefont {M.}~\bibnamefont {Dalmonte}},\ }\href {\doibase 10.1103/PhysRevB.102.014315} {\bibfield  {journal} {\bibinfo  {journal} {Phys. Rev. B}\ }\textbf {\bibinfo {volume} {102}},\ \bibinfo {pages} {014315} (\bibinfo {year} {2020})}\BibitemShut {NoStop}%
\bibitem [{\citenamefont {Li}\ \emph {et~al.}(2024)\citenamefont {Li}, \citenamefont {Vasseur}, \citenamefont {Fisher},\ and\ \citenamefont {Ludwig}}]{LiydPhysRevB109174307}%
  \BibitemOpen
  \bibfield  {author} {\bibinfo {author} {\bibfnamefont {Y.}~\bibnamefont {Li}}, \bibinfo {author} {\bibfnamefont {R.}~\bibnamefont {Vasseur}}, \bibinfo {author} {\bibfnamefont {M.~P.~A.}\ \bibnamefont {Fisher}}, \ and\ \bibinfo {author} {\bibfnamefont {A.~W.~W.}\ \bibnamefont {Ludwig}},\ }\href {\doibase 10.1103/PhysRevB.109.174307} {\bibfield  {journal} {\bibinfo  {journal} {Phys. Rev. B}\ }\textbf {\bibinfo {volume} {109}},\ \bibinfo {pages} {174307} (\bibinfo {year} {2024})}\BibitemShut {NoStop}%
\bibitem [{\citenamefont {Liu}\ \emph {et~al.}(2024{\natexlab{a}})\citenamefont {Liu}, \citenamefont {Li}, \citenamefont {Zhang}, \citenamefont {Jian},\ and\ \citenamefont {Yao}}]{LiusPhysRevB110064323}%
  \BibitemOpen
  \bibfield  {author} {\bibinfo {author} {\bibfnamefont {S.}~\bibnamefont {Liu}}, \bibinfo {author} {\bibfnamefont {M.-R.}\ \bibnamefont {Li}}, \bibinfo {author} {\bibfnamefont {S.-X.}\ \bibnamefont {Zhang}}, \bibinfo {author} {\bibfnamefont {S.-K.}\ \bibnamefont {Jian}}, \ and\ \bibinfo {author} {\bibfnamefont {H.}~\bibnamefont {Yao}},\ }\href {\doibase 10.1103/PhysRevB.110.064323} {\bibfield  {journal} {\bibinfo  {journal} {Phys. Rev. B}\ }\textbf {\bibinfo {volume} {110}},\ \bibinfo {pages} {064323} (\bibinfo {year} {2024}{\natexlab{a}})}\BibitemShut {NoStop}%
\bibitem [{\citenamefont {Liu}\ \emph {et~al.}(2024{\natexlab{b}})\citenamefont {Liu}, \citenamefont {Li}, \citenamefont {Zhang},\ and\ \citenamefont {Jian}}]{LiusPhysRevLett132240402}%
  \BibitemOpen
  \bibfield  {author} {\bibinfo {author} {\bibfnamefont {S.}~\bibnamefont {Liu}}, \bibinfo {author} {\bibfnamefont {M.-R.}\ \bibnamefont {Li}}, \bibinfo {author} {\bibfnamefont {S.-X.}\ \bibnamefont {Zhang}}, \ and\ \bibinfo {author} {\bibfnamefont {S.-K.}\ \bibnamefont {Jian}},\ }\href {\doibase 10.1103/PhysRevLett.132.240402} {\bibfield  {journal} {\bibinfo  {journal} {Phys. Rev. Lett.}\ }\textbf {\bibinfo {volume} {132}},\ \bibinfo {pages} {240402} (\bibinfo {year} {2024}{\natexlab{b}})}\BibitemShut {NoStop}%
\bibitem [{\citenamefont {Cecile}\ \emph {et~al.}(2024)\citenamefont {Cecile}, \citenamefont {L\'oio},\ and\ \citenamefont {De~Nardis}}]{CecilePhysRevResearch6033220}%
  \BibitemOpen
  \bibfield  {author} {\bibinfo {author} {\bibfnamefont {G.}~\bibnamefont {Cecile}}, \bibinfo {author} {\bibfnamefont {H.}~\bibnamefont {L\'oio}}, \ and\ \bibinfo {author} {\bibfnamefont {J.}~\bibnamefont {De~Nardis}},\ }\href {\doibase 10.1103/PhysRevResearch.6.033220} {\bibfield  {journal} {\bibinfo  {journal} {Phys. Rev. Res.}\ }\textbf {\bibinfo {volume} {6}},\ \bibinfo {pages} {033220} (\bibinfo {year} {2024})}\BibitemShut {NoStop}%
\bibitem [{\citenamefont {Lee}\ \emph {et~al.}(2023)\citenamefont {Lee}, \citenamefont {Jian},\ and\ \citenamefont {Xu}}]{XuckPRXQuantum4030317}%
  \BibitemOpen
  \bibfield  {author} {\bibinfo {author} {\bibfnamefont {J.~Y.}\ \bibnamefont {Lee}}, \bibinfo {author} {\bibfnamefont {C.-M.}\ \bibnamefont {Jian}}, \ and\ \bibinfo {author} {\bibfnamefont {C.}~\bibnamefont {Xu}},\ }\href {\doibase 10.1103/PRXQuantum.4.030317} {\bibfield  {journal} {\bibinfo  {journal} {PRX Quantum}\ }\textbf {\bibinfo {volume} {4}},\ \bibinfo {pages} {030317} (\bibinfo {year} {2023})}\BibitemShut {NoStop}%
\bibitem [{\citenamefont {Ashida}\ \emph {et~al.}(2024)\citenamefont {Ashida}, \citenamefont {Furukawa},\ and\ \citenamefont {Oshikawa}}]{AshidaPhysRevB110094404}%
  \BibitemOpen
  \bibfield  {author} {\bibinfo {author} {\bibfnamefont {Y.}~\bibnamefont {Ashida}}, \bibinfo {author} {\bibfnamefont {S.}~\bibnamefont {Furukawa}}, \ and\ \bibinfo {author} {\bibfnamefont {M.}~\bibnamefont {Oshikawa}},\ }\href {\doibase 10.1103/PhysRevB.110.094404} {\bibfield  {journal} {\bibinfo  {journal} {Phys. Rev. B}\ }\textbf {\bibinfo {volume} {110}},\ \bibinfo {pages} {094404} (\bibinfo {year} {2024})}\BibitemShut {NoStop}%
\bibitem [{\citenamefont {Murciano}\ \emph {et~al.}(2023)\citenamefont {Murciano}, \citenamefont {Sala}, \citenamefont {Liu}, \citenamefont {Mong},\ and\ \citenamefont {Alicea}}]{AliceaPhysRevX13041042}%
  \BibitemOpen
  \bibfield  {author} {\bibinfo {author} {\bibfnamefont {S.}~\bibnamefont {Murciano}}, \bibinfo {author} {\bibfnamefont {P.}~\bibnamefont {Sala}}, \bibinfo {author} {\bibfnamefont {Y.}~\bibnamefont {Liu}}, \bibinfo {author} {\bibfnamefont {R.~S.~K.}\ \bibnamefont {Mong}}, \ and\ \bibinfo {author} {\bibfnamefont {J.}~\bibnamefont {Alicea}},\ }\href {\doibase 10.1103/PhysRevX.13.041042} {\bibfield  {journal} {\bibinfo  {journal} {Phys. Rev. X}\ }\textbf {\bibinfo {volume} {13}},\ \bibinfo {pages} {041042} (\bibinfo {year} {2023})}\BibitemShut {NoStop}%
\bibitem [{\citenamefont {Moghaddam}\ \emph {et~al.}(2023)\citenamefont {Moghaddam}, \citenamefont {P\"oyh\"onen},\ and\ \citenamefont {Ojanen}}]{MoghaddamPhysRevLett131020401}%
  \BibitemOpen
  \bibfield  {author} {\bibinfo {author} {\bibfnamefont {A.~G.}\ \bibnamefont {Moghaddam}}, \bibinfo {author} {\bibfnamefont {K.}~\bibnamefont {P\"oyh\"onen}}, \ and\ \bibinfo {author} {\bibfnamefont {T.}~\bibnamefont {Ojanen}},\ }\href {\doibase 10.1103/PhysRevLett.131.020401} {\bibfield  {journal} {\bibinfo  {journal} {Phys. Rev. Lett.}\ }\textbf {\bibinfo {volume} {131}},\ \bibinfo {pages} {020401} (\bibinfo {year} {2023})}\BibitemShut {NoStop}%
\bibitem [{\citenamefont {Su}\ \emph {et~al.}(2024)\citenamefont {Su}, \citenamefont {Myerson-Jain},\ and\ \citenamefont {Xu}}]{XuckPhysRevB109035146}%
  \BibitemOpen
  \bibfield  {author} {\bibinfo {author} {\bibfnamefont {K.}~\bibnamefont {Su}}, \bibinfo {author} {\bibfnamefont {N.}~\bibnamefont {Myerson-Jain}}, \ and\ \bibinfo {author} {\bibfnamefont {C.}~\bibnamefont {Xu}},\ }\href {\doibase 10.1103/PhysRevB.109.035146} {\bibfield  {journal} {\bibinfo  {journal} {Phys. Rev. B}\ }\textbf {\bibinfo {volume} {109}},\ \bibinfo {pages} {035146} (\bibinfo {year} {2024})}\BibitemShut {NoStop}%
\bibitem [{\citenamefont {Li}\ \emph {et~al.}(2023)\citenamefont {Li}, \citenamefont {Zou}, \citenamefont {Glorioso}, \citenamefont {Altman},\ and\ \citenamefont {Fisher}}]{FisherPhysRevLett130220404}%
  \BibitemOpen
  \bibfield  {author} {\bibinfo {author} {\bibfnamefont {Y.}~\bibnamefont {Li}}, \bibinfo {author} {\bibfnamefont {Y.}~\bibnamefont {Zou}}, \bibinfo {author} {\bibfnamefont {P.}~\bibnamefont {Glorioso}}, \bibinfo {author} {\bibfnamefont {E.}~\bibnamefont {Altman}}, \ and\ \bibinfo {author} {\bibfnamefont {M.~P.~A.}\ \bibnamefont {Fisher}},\ }\href {\doibase 10.1103/PhysRevLett.130.220404} {\bibfield  {journal} {\bibinfo  {journal} {Phys. Rev. Lett.}\ }\textbf {\bibinfo {volume} {130}},\ \bibinfo {pages} {220404} (\bibinfo {year} {2023})}\BibitemShut {NoStop}%
\bibitem [{\citenamefont {Morral-Yepes}\ \emph {et~al.}(2024)\citenamefont {Morral-Yepes}, \citenamefont {Smith}, \citenamefont {Sondhi},\ and\ \citenamefont {Pollmann}}]{PolmannPRXQuantum5010309}%
  \BibitemOpen
  \bibfield  {author} {\bibinfo {author} {\bibfnamefont {R.}~\bibnamefont {Morral-Yepes}}, \bibinfo {author} {\bibfnamefont {A.}~\bibnamefont {Smith}}, \bibinfo {author} {\bibfnamefont {S.}~\bibnamefont {Sondhi}}, \ and\ \bibinfo {author} {\bibfnamefont {F.}~\bibnamefont {Pollmann}},\ }\href {\doibase 10.1103/PRXQuantum.5.010309} {\bibfield  {journal} {\bibinfo  {journal} {PRX Quantum}\ }\textbf {\bibinfo {volume} {5}},\ \bibinfo {pages} {010309} (\bibinfo {year} {2024})}\BibitemShut {NoStop}%
\bibitem [{\citenamefont {Lavasani}\ \emph {et~al.}(2021)\citenamefont {Lavasani}, \citenamefont {Alavirad},\ and\ \citenamefont {Barkeshli}}]{Lavasaninpexp}%
  \BibitemOpen
  \bibfield  {author} {\bibinfo {author} {\bibfnamefont {A.}~\bibnamefont {Lavasani}}, \bibinfo {author} {\bibfnamefont {Y.}~\bibnamefont {Alavirad}}, \ and\ \bibinfo {author} {\bibfnamefont {M.}~\bibnamefont {Barkeshli}},\ }\href {\doibase 10.1038/s41567-020-01112-z} {\bibfield  {journal} {\bibinfo  {journal} {Nature Physics}\ }\textbf {\bibinfo {volume} {17}},\ \bibinfo {pages} {242} (\bibinfo {year} {2021})}\BibitemShut {NoStop}%
\bibitem [{\citenamefont {Noel}\ \emph {et~al.}(2022)\citenamefont {Noel}, \citenamefont {Niroula}, \citenamefont {Zhu}, \citenamefont {Risinger}, \citenamefont {Egan}, \citenamefont {Biswas}, \citenamefont {Cetina}, \citenamefont {Gorshkov}, \citenamefont {Gullans}, \citenamefont {Huse},\ and\ \citenamefont {Monroe}}]{Noelnpexp}%
  \BibitemOpen
  \bibfield  {author} {\bibinfo {author} {\bibfnamefont {C.}~\bibnamefont {Noel}}, \bibinfo {author} {\bibfnamefont {P.}~\bibnamefont {Niroula}}, \bibinfo {author} {\bibfnamefont {D.}~\bibnamefont {Zhu}}, \bibinfo {author} {\bibfnamefont {A.}~\bibnamefont {Risinger}}, \bibinfo {author} {\bibfnamefont {L.}~\bibnamefont {Egan}}, \bibinfo {author} {\bibfnamefont {D.}~\bibnamefont {Biswas}}, \bibinfo {author} {\bibfnamefont {M.}~\bibnamefont {Cetina}}, \bibinfo {author} {\bibfnamefont {A.~V.}\ \bibnamefont {Gorshkov}}, \bibinfo {author} {\bibfnamefont {M.~J.}\ \bibnamefont {Gullans}}, \bibinfo {author} {\bibfnamefont {D.~A.}\ \bibnamefont {Huse}}, \ and\ \bibinfo {author} {\bibfnamefont {C.}~\bibnamefont {Monroe}},\ }\href {\doibase 10.1038/s41567-022-01619-7} {\bibfield  {journal} {\bibinfo  {journal} {Nature Physics}\ }\textbf {\bibinfo {volume} {18}},\ \bibinfo {pages} {760} (\bibinfo {year} {2022})}\BibitemShut {NoStop}%
\bibitem [{\citenamefont {Koh}\ \emph {et~al.}(2023)\citenamefont {Koh}, \citenamefont {Sun}, \citenamefont {Motta},\ and\ \citenamefont {Minnich}}]{Koh2023}%
  \BibitemOpen
  \bibfield  {author} {\bibinfo {author} {\bibfnamefont {J.~M.}\ \bibnamefont {Koh}}, \bibinfo {author} {\bibfnamefont {S.-N.}\ \bibnamefont {Sun}}, \bibinfo {author} {\bibfnamefont {M.}~\bibnamefont {Motta}}, \ and\ \bibinfo {author} {\bibfnamefont {A.~J.}\ \bibnamefont {Minnich}},\ }\href {\doibase 10.1038/s41567-023-02076-6} {\bibfield  {journal} {\bibinfo  {journal} {Nature Physics}\ }\textbf {\bibinfo {volume} {19}},\ \bibinfo {pages} {1314} (\bibinfo {year} {2023})}\BibitemShut {NoStop}%
\bibitem [{\citenamefont {AI}\ and\ \citenamefont {Collaborators}(2023)}]{Hoke2023}%
  \BibitemOpen
  \bibfield  {author} {\bibinfo {author} {\bibfnamefont {G.~Q.}\ \bibnamefont {AI}}\ and\ \bibinfo {author} {\bibnamefont {Collaborators}},\ }\href {\doibase 10.1038/s41586-023-06505-7} {\bibfield  {journal} {\bibinfo  {journal} {Nature}\ }\textbf {\bibinfo {volume} {622}},\ \bibinfo {pages} {481} (\bibinfo {year} {2023})}\BibitemShut {NoStop}%
\bibitem [{\citenamefont {Weinstein}\ \emph {et~al.}(2022)\citenamefont {Weinstein}, \citenamefont {Bao},\ and\ \citenamefont {Altman}}]{PhysRevLett.129.080501}%
  \BibitemOpen
  \bibfield  {author} {\bibinfo {author} {\bibfnamefont {Z.}~\bibnamefont {Weinstein}}, \bibinfo {author} {\bibfnamefont {Y.}~\bibnamefont {Bao}}, \ and\ \bibinfo {author} {\bibfnamefont {E.}~\bibnamefont {Altman}},\ }\href {\doibase 10.1103/PhysRevLett.129.080501} {\bibfield  {journal} {\bibinfo  {journal} {Phys. Rev. Lett.}\ }\textbf {\bibinfo {volume} {129}},\ \bibinfo {pages} {080501} (\bibinfo {year} {2022})}\BibitemShut {NoStop}%
\bibitem [{\citenamefont {Sang}\ and\ \citenamefont {Hsieh}(2021)}]{PhysRevResearch.3.023200}%
  \BibitemOpen
  \bibfield  {author} {\bibinfo {author} {\bibfnamefont {S.}~\bibnamefont {Sang}}\ and\ \bibinfo {author} {\bibfnamefont {T.~H.}\ \bibnamefont {Hsieh}},\ }\href {\doibase 10.1103/PhysRevResearch.3.023200} {\bibfield  {journal} {\bibinfo  {journal} {Phys. Rev. Res.}\ }\textbf {\bibinfo {volume} {3}},\ \bibinfo {pages} {023200} (\bibinfo {year} {2021})}\BibitemShut {NoStop}%
\bibitem [{\citenamefont {Kelly}\ and\ \citenamefont {Marino}(2024)}]{kelly2024generalizingmeasurementinducedphasetransitions}%
  \BibitemOpen
  \bibfield  {author} {\bibinfo {author} {\bibfnamefont {S.~P.}\ \bibnamefont {Kelly}}\ and\ \bibinfo {author} {\bibfnamefont {J.}~\bibnamefont {Marino}},\ }\href {https://arxiv.org/abs/2402.13271} {\bibfield  {journal} {\bibinfo  {journal} {arXiv:2402.13271}\ } (\bibinfo {year} {2024})}\BibitemShut {NoStop}%
\bibitem [{\citenamefont {Chen}\ \emph {et~al.}(2024)\citenamefont {Chen}, \citenamefont {Liu},\ and\ \citenamefont {Zhang}}]{Chen2024_z}%
  \BibitemOpen
  \bibfield  {author} {\bibinfo {author} {\bibfnamefont {Y.-Q.}\ \bibnamefont {Chen}}, \bibinfo {author} {\bibfnamefont {S.}~\bibnamefont {Liu}}, \ and\ \bibinfo {author} {\bibfnamefont {S.-X.}\ \bibnamefont {Zhang}},\ }\href {https://arxiv.org/abs/2405.05076} {\bibfield  {journal} {\bibinfo  {journal} {arXiv:2405.05076}\ } (\bibinfo {year} {2024})}\BibitemShut {NoStop}%
\bibitem [{\citenamefont {Qian}\ and\ \citenamefont {Wang}(2024)}]{qian2024coherentinformationphasetransition}%
  \BibitemOpen
  \bibfield  {author} {\bibinfo {author} {\bibfnamefont {D.}~\bibnamefont {Qian}}\ and\ \bibinfo {author} {\bibfnamefont {J.}~\bibnamefont {Wang}},\ }\href {https://arxiv.org/abs/2408.16267} {\bibfield  {journal} {\bibinfo  {journal} {arXiv:2408.16267}\ } (\bibinfo {year} {2024})}\BibitemShut {NoStop}%
\bibitem [{\citenamefont {Feng}\ \emph {et~al.}(2023)\citenamefont {Feng}, \citenamefont {Liu}, \citenamefont {Chen},\ and\ \citenamefont {Guo}}]{PhysRevB.107.094309}%
  \BibitemOpen
  \bibfield  {author} {\bibinfo {author} {\bibfnamefont {X.}~\bibnamefont {Feng}}, \bibinfo {author} {\bibfnamefont {S.}~\bibnamefont {Liu}}, \bibinfo {author} {\bibfnamefont {S.}~\bibnamefont {Chen}}, \ and\ \bibinfo {author} {\bibfnamefont {W.}~\bibnamefont {Guo}},\ }\href {\doibase 10.1103/PhysRevB.107.094309} {\bibfield  {journal} {\bibinfo  {journal} {Phys. Rev. B}\ }\textbf {\bibinfo {volume} {107}},\ \bibinfo {pages} {094309} (\bibinfo {year} {2023})}\BibitemShut {NoStop}%
\bibitem [{\citenamefont {Yu}\ \emph {et~al.}(2025)\citenamefont {Yu}, \citenamefont {Yang}, \citenamefont {Liu}, \citenamefont {Lin},\ and\ \citenamefont {Jian}}]{yu2025gapless}%
  \BibitemOpen
  \bibfield  {author} {\bibinfo {author} {\bibfnamefont {X.-J.}\ \bibnamefont {Yu}}, \bibinfo {author} {\bibfnamefont {S.}~\bibnamefont {Yang}}, \bibinfo {author} {\bibfnamefont {S.}~\bibnamefont {Liu}}, \bibinfo {author} {\bibfnamefont {H.-Q.}\ \bibnamefont {Lin}}, \ and\ \bibinfo {author} {\bibfnamefont {S.-K.}\ \bibnamefont {Jian}},\ }\href {https://arxiv.org/abs/2501.03851} {\bibfield  {journal} {\bibinfo  {journal} {arXiv:2501.03851}\ } (\bibinfo {year} {2025})}\BibitemShut {NoStop}%
\bibitem [{\citenamefont {Hohenberg}\ and\ \citenamefont {Halperin}(1977)}]{Hohenberg1977rmp}%
  \BibitemOpen
  \bibfield  {author} {\bibinfo {author} {\bibfnamefont {P.~C.}\ \bibnamefont {Hohenberg}}\ and\ \bibinfo {author} {\bibfnamefont {B.~I.}\ \bibnamefont {Halperin}},\ }\href {\doibase 10.1103/RevModPhys.49.435} {\bibfield  {journal} {\bibinfo  {journal} {Rev. Mod. Phys.}\ }\textbf {\bibinfo {volume} {49}},\ \bibinfo {pages} {435} (\bibinfo {year} {1977})}\BibitemShut {NoStop}%
\bibitem [{\citenamefont {Janssen}\ \emph {et~al.}(1989)\citenamefont {Janssen}, \citenamefont {Schaub},\ and\ \citenamefont {Schmittmann}}]{Janssen1989}%
  \BibitemOpen
  \bibfield  {author} {\bibinfo {author} {\bibfnamefont {H.~K.}\ \bibnamefont {Janssen}}, \bibinfo {author} {\bibfnamefont {B.}~\bibnamefont {Schaub}}, \ and\ \bibinfo {author} {\bibfnamefont {B.}~\bibnamefont {Schmittmann}},\ }\href {\doibase 10.1007/BF01319383} {\bibfield  {journal} {\bibinfo  {journal} {Zeitschrift f\"{u}r Physik B Condensed Matter}\ }\textbf {\bibinfo {volume} {73}},\ \bibinfo {pages} {539} (\bibinfo {year} {1989})}\BibitemShut {NoStop}%
\bibitem [{\citenamefont {Li}\ \emph {et~al.}(1995)\citenamefont {Li}, \citenamefont {Sch\"ulke},\ and\ \citenamefont {Zheng}}]{Lizb1995prl}%
  \BibitemOpen
  \bibfield  {author} {\bibinfo {author} {\bibfnamefont {Z.~B.}\ \bibnamefont {Li}}, \bibinfo {author} {\bibfnamefont {L.}~\bibnamefont {Sch\"ulke}}, \ and\ \bibinfo {author} {\bibfnamefont {B.}~\bibnamefont {Zheng}},\ }\href {\doibase 10.1103/PhysRevLett.74.3396} {\bibfield  {journal} {\bibinfo  {journal} {Phys. Rev. Lett.}\ }\textbf {\bibinfo {volume} {74}},\ \bibinfo {pages} {3396} (\bibinfo {year} {1995})}\BibitemShut {NoStop}%
\bibitem [{\citenamefont {Zheng}(1996)}]{Zhengb1996prl}%
  \BibitemOpen
  \bibfield  {author} {\bibinfo {author} {\bibfnamefont {B.}~\bibnamefont {Zheng}},\ }\href {\doibase 10.1103/PhysRevLett.77.679} {\bibfield  {journal} {\bibinfo  {journal} {Phys. Rev. Lett.}\ }\textbf {\bibinfo {volume} {77}},\ \bibinfo {pages} {679} (\bibinfo {year} {1996})}\BibitemShut {NoStop}%
\bibitem [{\citenamefont {Zheng}(1998)}]{Zhengb1998review}%
  \BibitemOpen
  \bibfield  {author} {\bibinfo {author} {\bibfnamefont {B.}~\bibnamefont {Zheng}},\ }\href {\doibase 10.1142/S021797929800288X} {\bibfield  {journal} {\bibinfo  {journal} {International Journal of Modern Physics B}\ }\textbf {\bibinfo {volume} {12}},\ \bibinfo {pages} {1419} (\bibinfo {year} {1998})},\ \Eprint {http://arxiv.org/abs/https://doi.org/10.1142/S021797929800288X} {https://doi.org/10.1142/S021797929800288X} \BibitemShut {NoStop}%
\bibitem [{\citenamefont {Albano}\ \emph {et~al.}(2011)\citenamefont {Albano}, \citenamefont {Bab}, \citenamefont {Baglietto}, \citenamefont {Borzi}, \citenamefont {Grigera}, \citenamefont {Loscar}, \citenamefont {Rodriguez}, \citenamefont {Puzzo},\ and\ \citenamefont {Saracco}}]{Albano2011iop}%
  \BibitemOpen
  \bibfield  {author} {\bibinfo {author} {\bibfnamefont {E.~V.}\ \bibnamefont {Albano}}, \bibinfo {author} {\bibfnamefont {M.~A.}\ \bibnamefont {Bab}}, \bibinfo {author} {\bibfnamefont {G.}~\bibnamefont {Baglietto}}, \bibinfo {author} {\bibfnamefont {R.~A.}\ \bibnamefont {Borzi}}, \bibinfo {author} {\bibfnamefont {T.~S.}\ \bibnamefont {Grigera}}, \bibinfo {author} {\bibfnamefont {E.~S.}\ \bibnamefont {Loscar}}, \bibinfo {author} {\bibfnamefont {D.~E.}\ \bibnamefont {Rodriguez}}, \bibinfo {author} {\bibfnamefont {M.~L.~R.}\ \bibnamefont {Puzzo}}, \ and\ \bibinfo {author} {\bibfnamefont {G.~P.}\ \bibnamefont {Saracco}},\ }\href {\doibase 10.1088/0034-4885/74/2/026501} {\bibfield  {journal} {\bibinfo  {journal} {Reports on Progress in Physics}\ }\textbf {\bibinfo {volume} {74}},\ \bibinfo {pages} {026501} (\bibinfo {year} {2011})}\BibitemShut {NoStop}%
\bibitem [{\citenamefont {Shu}\ \emph {et~al.}(2024)\citenamefont {Shu}, \citenamefont {Liao},\ and\ \citenamefont {Yin}}]{Shuyr2024prb}%
  \BibitemOpen
  \bibfield  {author} {\bibinfo {author} {\bibfnamefont {Y.-R.}\ \bibnamefont {Shu}}, \bibinfo {author} {\bibfnamefont {T.}~\bibnamefont {Liao}}, \ and\ \bibinfo {author} {\bibfnamefont {S.}~\bibnamefont {Yin}},\ }\href {\doibase 10.1103/PhysRevB.110.134306} {\bibfield  {journal} {\bibinfo  {journal} {Phys. Rev. B}\ }\textbf {\bibinfo {volume} {110}},\ \bibinfo {pages} {134306} (\bibinfo {year} {2024})}\BibitemShut {NoStop}%
\bibitem [{\citenamefont {Dalla~Torre}\ \emph {et~al.}(2013)\citenamefont {Dalla~Torre}, \citenamefont {Demler},\ and\ \citenamefont {Polkovnikov}}]{Polkovnikov2013prl}%
  \BibitemOpen
  \bibfield  {author} {\bibinfo {author} {\bibfnamefont {E.~G.}\ \bibnamefont {Dalla~Torre}}, \bibinfo {author} {\bibfnamefont {E.}~\bibnamefont {Demler}}, \ and\ \bibinfo {author} {\bibfnamefont {A.}~\bibnamefont {Polkovnikov}},\ }\href {\doibase 10.1103/PhysRevLett.110.090404} {\bibfield  {journal} {\bibinfo  {journal} {Phys. Rev. Lett.}\ }\textbf {\bibinfo {volume} {110}},\ \bibinfo {pages} {090404} (\bibinfo {year} {2013})}\BibitemShut {NoStop}%
\bibitem [{\citenamefont {Gagel}\ \emph {et~al.}(2014)\citenamefont {Gagel}, \citenamefont {Orth},\ and\ \citenamefont {Schmalian}}]{Schmalian2014prl}%
  \BibitemOpen
  \bibfield  {author} {\bibinfo {author} {\bibfnamefont {P.}~\bibnamefont {Gagel}}, \bibinfo {author} {\bibfnamefont {P.~P.}\ \bibnamefont {Orth}}, \ and\ \bibinfo {author} {\bibfnamefont {J.}~\bibnamefont {Schmalian}},\ }\href {\doibase 10.1103/PhysRevLett.113.220401} {\bibfield  {journal} {\bibinfo  {journal} {Phys. Rev. Lett.}\ }\textbf {\bibinfo {volume} {113}},\ \bibinfo {pages} {220401} (\bibinfo {year} {2014})}\BibitemShut {NoStop}%
\bibitem [{\citenamefont {Gagel}\ \emph {et~al.}(2015)\citenamefont {Gagel}, \citenamefont {Orth},\ and\ \citenamefont {Schmalian}}]{Schmalian2015prb}%
  \BibitemOpen
  \bibfield  {author} {\bibinfo {author} {\bibfnamefont {P.}~\bibnamefont {Gagel}}, \bibinfo {author} {\bibfnamefont {P.~P.}\ \bibnamefont {Orth}}, \ and\ \bibinfo {author} {\bibfnamefont {J.}~\bibnamefont {Schmalian}},\ }\href {\doibase 10.1103/PhysRevB.92.115121} {\bibfield  {journal} {\bibinfo  {journal} {Phys. Rev. B}\ }\textbf {\bibinfo {volume} {92}},\ \bibinfo {pages} {115121} (\bibinfo {year} {2015})}\BibitemShut {NoStop}%
\bibitem [{\citenamefont {Chiocchetta}\ \emph {et~al.}(2015)\citenamefont {Chiocchetta}, \citenamefont {Tavora}, \citenamefont {Gambassi},\ and\ \citenamefont {Mitra}}]{Chiocchetta2015prb}%
  \BibitemOpen
  \bibfield  {author} {\bibinfo {author} {\bibfnamefont {A.}~\bibnamefont {Chiocchetta}}, \bibinfo {author} {\bibfnamefont {M.}~\bibnamefont {Tavora}}, \bibinfo {author} {\bibfnamefont {A.}~\bibnamefont {Gambassi}}, \ and\ \bibinfo {author} {\bibfnamefont {A.}~\bibnamefont {Mitra}},\ }\href {\doibase 10.1103/PhysRevB.91.220302} {\bibfield  {journal} {\bibinfo  {journal} {Phys. Rev. B}\ }\textbf {\bibinfo {volume} {91}},\ \bibinfo {pages} {220302} (\bibinfo {year} {2015})}\BibitemShut {NoStop}%
\bibitem [{\citenamefont {Chiocchetta}\ \emph {et~al.}(2017)\citenamefont {Chiocchetta}, \citenamefont {Gambassi}, \citenamefont {Diehl},\ and\ \citenamefont {Marino}}]{Marino201prl}%
  \BibitemOpen
  \bibfield  {author} {\bibinfo {author} {\bibfnamefont {A.}~\bibnamefont {Chiocchetta}}, \bibinfo {author} {\bibfnamefont {A.}~\bibnamefont {Gambassi}}, \bibinfo {author} {\bibfnamefont {S.}~\bibnamefont {Diehl}}, \ and\ \bibinfo {author} {\bibfnamefont {J.}~\bibnamefont {Marino}},\ }\href {\doibase 10.1103/PhysRevLett.118.135701} {\bibfield  {journal} {\bibinfo  {journal} {Phys. Rev. Lett.}\ }\textbf {\bibinfo {volume} {118}},\ \bibinfo {pages} {135701} (\bibinfo {year} {2017})}\BibitemShut {NoStop}%
\bibitem [{\citenamefont {Jian}\ \emph {et~al.}(2019)\citenamefont {Jian}, \citenamefont {Yin},\ and\ \citenamefont {Swingle}}]{Yin2019prl}%
  \BibitemOpen
  \bibfield  {author} {\bibinfo {author} {\bibfnamefont {S.-K.}\ \bibnamefont {Jian}}, \bibinfo {author} {\bibfnamefont {S.}~\bibnamefont {Yin}}, \ and\ \bibinfo {author} {\bibfnamefont {B.}~\bibnamefont {Swingle}},\ }\href {\doibase 10.1103/PhysRevLett.123.170606} {\bibfield  {journal} {\bibinfo  {journal} {Phys. Rev. Lett.}\ }\textbf {\bibinfo {volume} {123}},\ \bibinfo {pages} {170606} (\bibinfo {year} {2019})}\BibitemShut {NoStop}%
\bibitem [{\citenamefont {Yin}\ and\ \citenamefont {Jian}(2021)}]{Yin2021prb}%
  \BibitemOpen
  \bibfield  {author} {\bibinfo {author} {\bibfnamefont {S.}~\bibnamefont {Yin}}\ and\ \bibinfo {author} {\bibfnamefont {S.-K.}\ \bibnamefont {Jian}},\ }\href {\doibase 10.1103/PhysRevB.103.125116} {\bibfield  {journal} {\bibinfo  {journal} {Phys. Rev. B}\ }\textbf {\bibinfo {volume} {103}},\ \bibinfo {pages} {125116} (\bibinfo {year} {2021})}\BibitemShut {NoStop}%
\bibitem [{\citenamefont {Marino}\ \emph {et~al.}(2022)\citenamefont {Marino}, \citenamefont {Eckstein}, \citenamefont {Foster},\ and\ \citenamefont {Rey}}]{Marino_2022}%
  \BibitemOpen
  \bibfield  {author} {\bibinfo {author} {\bibfnamefont {J.}~\bibnamefont {Marino}}, \bibinfo {author} {\bibfnamefont {M.}~\bibnamefont {Eckstein}}, \bibinfo {author} {\bibfnamefont {M.~S.}\ \bibnamefont {Foster}}, \ and\ \bibinfo {author} {\bibfnamefont {A.~M.}\ \bibnamefont {Rey}},\ }\href {\doibase 10.1088/1361-6633/ac906c} {\bibfield  {journal} {\bibinfo  {journal} {Reports on Progress in Physics}\ }\textbf {\bibinfo {volume} {85}},\ \bibinfo {pages} {116001} (\bibinfo {year} {2022})}\BibitemShut {NoStop}%
\bibitem [{\citenamefont {Yin}\ \emph {et~al.}(2014)\citenamefont {Yin}, \citenamefont {Mai},\ and\ \citenamefont {Zhong}}]{Yins2014prb}%
  \BibitemOpen
  \bibfield  {author} {\bibinfo {author} {\bibfnamefont {S.}~\bibnamefont {Yin}}, \bibinfo {author} {\bibfnamefont {P.}~\bibnamefont {Mai}}, \ and\ \bibinfo {author} {\bibfnamefont {F.}~\bibnamefont {Zhong}},\ }\href {\doibase 10.1103/PhysRevB.89.144115} {\bibfield  {journal} {\bibinfo  {journal} {Phys. Rev. B}\ }\textbf {\bibinfo {volume} {89}},\ \bibinfo {pages} {144115} (\bibinfo {year} {2014})}\BibitemShut {NoStop}%
\bibitem [{\citenamefont {Zhang}\ \emph {et~al.}(2014)\citenamefont {Zhang}, \citenamefont {Yin},\ and\ \citenamefont {Zhong}}]{Yins2014pre}%
  \BibitemOpen
  \bibfield  {author} {\bibinfo {author} {\bibfnamefont {S.}~\bibnamefont {Zhang}}, \bibinfo {author} {\bibfnamefont {S.}~\bibnamefont {Yin}}, \ and\ \bibinfo {author} {\bibfnamefont {F.}~\bibnamefont {Zhong}},\ }\href {\doibase 10.1103/PhysRevE.90.042104} {\bibfield  {journal} {\bibinfo  {journal} {Phys. Rev. E}\ }\textbf {\bibinfo {volume} {90}},\ \bibinfo {pages} {042104} (\bibinfo {year} {2014})}\BibitemShut {NoStop}%
\bibitem [{\citenamefont {Shu}\ \emph {et~al.}(2017)\citenamefont {Shu}, \citenamefont {Yin},\ and\ \citenamefont {Yao}}]{Shu2017prb}%
  \BibitemOpen
  \bibfield  {author} {\bibinfo {author} {\bibfnamefont {Y.-R.}\ \bibnamefont {Shu}}, \bibinfo {author} {\bibfnamefont {S.}~\bibnamefont {Yin}}, \ and\ \bibinfo {author} {\bibfnamefont {D.-X.}\ \bibnamefont {Yao}},\ }\href {\doibase 10.1103/PhysRevB.96.094304} {\bibfield  {journal} {\bibinfo  {journal} {Phys. Rev. B}\ }\textbf {\bibinfo {volume} {96}},\ \bibinfo {pages} {094304} (\bibinfo {year} {2017})}\BibitemShut {NoStop}%
\bibitem [{\citenamefont {Shu}\ \emph {et~al.}(2022)\citenamefont {Shu}, \citenamefont {Jian},\ and\ \citenamefont {Yin}}]{Yin2022prl}%
  \BibitemOpen
  \bibfield  {author} {\bibinfo {author} {\bibfnamefont {Y.-R.}\ \bibnamefont {Shu}}, \bibinfo {author} {\bibfnamefont {S.-K.}\ \bibnamefont {Jian}}, \ and\ \bibinfo {author} {\bibfnamefont {S.}~\bibnamefont {Yin}},\ }\href {\doibase 10.1103/PhysRevLett.128.020601} {\bibfield  {journal} {\bibinfo  {journal} {Phys. Rev. Lett.}\ }\textbf {\bibinfo {volume} {128}},\ \bibinfo {pages} {020601} (\bibinfo {year} {2022})}\BibitemShut {NoStop}%
\bibitem [{\citenamefont {Shu}\ and\ \citenamefont {Yin}(2022)}]{Yin2022prb}%
  \BibitemOpen
  \bibfield  {author} {\bibinfo {author} {\bibfnamefont {Y.-R.}\ \bibnamefont {Shu}}\ and\ \bibinfo {author} {\bibfnamefont {S.}~\bibnamefont {Yin}},\ }\href {\doibase 10.1103/PhysRevB.105.104420} {\bibfield  {journal} {\bibinfo  {journal} {Phys. Rev. B}\ }\textbf {\bibinfo {volume} {105}},\ \bibinfo {pages} {104420} (\bibinfo {year} {2022})}\BibitemShut {NoStop}%
\bibitem [{\citenamefont {Zuo}\ \emph {et~al.}(2021)\citenamefont {Zuo}, \citenamefont {Yin}, \citenamefont {Cao},\ and\ \citenamefont {Zhong}}]{Zuo2021prb}%
  \BibitemOpen
  \bibfield  {author} {\bibinfo {author} {\bibfnamefont {Z.}~\bibnamefont {Zuo}}, \bibinfo {author} {\bibfnamefont {S.}~\bibnamefont {Yin}}, \bibinfo {author} {\bibfnamefont {X.}~\bibnamefont {Cao}}, \ and\ \bibinfo {author} {\bibfnamefont {F.}~\bibnamefont {Zhong}},\ }\href {\doibase 10.1103/PhysRevB.104.214108} {\bibfield  {journal} {\bibinfo  {journal} {Phys. Rev. B}\ }\textbf {\bibinfo {volume} {104}},\ \bibinfo {pages} {214108} (\bibinfo {year} {2021})}\BibitemShut {NoStop}%
\bibitem [{\citenamefont {Cai}\ \emph {et~al.}(2024)\citenamefont {Cai}, \citenamefont {Shu}, \citenamefont {Rao},\ and\ \citenamefont {Yin}}]{Yin2024prbcor}%
  \BibitemOpen
  \bibfield  {author} {\bibinfo {author} {\bibfnamefont {J.-Q.}\ \bibnamefont {Cai}}, \bibinfo {author} {\bibfnamefont {Y.-R.}\ \bibnamefont {Shu}}, \bibinfo {author} {\bibfnamefont {X.-Q.}\ \bibnamefont {Rao}}, \ and\ \bibinfo {author} {\bibfnamefont {S.}~\bibnamefont {Yin}},\ }\href {\doibase 10.1103/PhysRevB.109.184303} {\bibfield  {journal} {\bibinfo  {journal} {Phys. Rev. B}\ }\textbf {\bibinfo {volume} {109}},\ \bibinfo {pages} {184303} (\bibinfo {year} {2024})}\BibitemShut {NoStop}%
\bibitem [{\citenamefont {Chang}\ \emph {et~al.}(2024)\citenamefont {Chang}, \citenamefont {Yin}, \citenamefont {Zhang},\ and\ \citenamefont {Li}}]{Changwx2024}%
  \BibitemOpen
  \bibfield  {author} {\bibinfo {author} {\bibfnamefont {W.-X.}\ \bibnamefont {Chang}}, \bibinfo {author} {\bibfnamefont {S.}~\bibnamefont {Yin}}, \bibinfo {author} {\bibfnamefont {S.-X.}\ \bibnamefont {Zhang}}, \ and\ \bibinfo {author} {\bibfnamefont {Z.-X.}\ \bibnamefont {Li}},\ }\href {\doibase 10.48550/arXiv.2409.06547} {\bibfield  {journal} {\bibinfo  {journal} {arXiv:2409.06547}\ } (\bibinfo {year} {2024}),\ 10.48550/arXiv.2409.06547}\BibitemShut {NoStop}%
\bibitem [{\citenamefont {Yu}\ \emph {et~al.}(2023)\citenamefont {Yu}, \citenamefont {Zeng}, \citenamefont {Shu}, \citenamefont {Li},\ and\ \citenamefont {Yin}}]{Yuarxiv2023}%
  \BibitemOpen
  \bibfield  {author} {\bibinfo {author} {\bibfnamefont {Y.-K.}\ \bibnamefont {Yu}}, \bibinfo {author} {\bibfnamefont {Z.}~\bibnamefont {Zeng}}, \bibinfo {author} {\bibfnamefont {Y.-R.}\ \bibnamefont {Shu}}, \bibinfo {author} {\bibfnamefont {Z.-X.}\ \bibnamefont {Li}}, \ and\ \bibinfo {author} {\bibfnamefont {S.}~\bibnamefont {Yin}},\ }\href {\doibase 10.48550/arXiv.2310.10601} {\bibfield  {journal} {\bibinfo  {journal} {arXiv: 2310.10601}\ } (\bibinfo {year} {2023}),\ 10.48550/arXiv.2310.10601}\BibitemShut {NoStop}%
\bibitem [{\citenamefont {Yu}\ \emph {et~al.}(2024)\citenamefont {Yu}, \citenamefont {Li}, \citenamefont {Yin},\ and\ \citenamefont {Li}}]{Yuarxiv2024}%
  \BibitemOpen
  \bibfield  {author} {\bibinfo {author} {\bibfnamefont {Y.-K.}\ \bibnamefont {Yu}}, \bibinfo {author} {\bibfnamefont {Z.-X.}\ \bibnamefont {Li}}, \bibinfo {author} {\bibfnamefont {S.}~\bibnamefont {Yin}}, \ and\ \bibinfo {author} {\bibfnamefont {Z.-X.}\ \bibnamefont {Li}},\ }\href {\doibase 10.48550/arXiv.2410.18854} {\bibfield  {journal} {\bibinfo  {journal} {arXiv: 2410.18854}\ } (\bibinfo {year} {2024}),\ 10.48550/arXiv.2410.18854}\BibitemShut {NoStop}%
\bibitem [{\citenamefont {Zhang}\ and\ \citenamefont {Yin}(2024)}]{Zhangsx2024prbys}%
  \BibitemOpen
  \bibfield  {author} {\bibinfo {author} {\bibfnamefont {S.-X.}\ \bibnamefont {Zhang}}\ and\ \bibinfo {author} {\bibfnamefont {S.}~\bibnamefont {Yin}},\ }\href {\doibase 10.1103/PhysRevB.109.134309} {\bibfield  {journal} {\bibinfo  {journal} {Phys. Rev. B}\ }\textbf {\bibinfo {volume} {109}},\ \bibinfo {pages} {134309} (\bibinfo {year} {2024})}\BibitemShut {NoStop}%
\bibitem [{\citenamefont {Wang}\ \emph {et~al.}(2024)\citenamefont {Wang}, \citenamefont {Liu}, \citenamefont {Li}, \citenamefont {Zhang},\ and\ \citenamefont {Yin}}]{Yinarxiv2024}%
  \BibitemOpen
  \bibfield  {author} {\bibinfo {author} {\bibfnamefont {W.}~\bibnamefont {Wang}}, \bibinfo {author} {\bibfnamefont {S.}~\bibnamefont {Liu}}, \bibinfo {author} {\bibfnamefont {J.}~\bibnamefont {Li}}, \bibinfo {author} {\bibfnamefont {S.-X.}\ \bibnamefont {Zhang}}, \ and\ \bibinfo {author} {\bibfnamefont {S.}~\bibnamefont {Yin}},\ }\href {\doibase 10.48550/arXiv.2411.06648} {\bibfield  {journal} {\bibinfo  {journal} {arXiv: 2411.06648}\ } (\bibinfo {year} {2024}),\ 10.48550/arXiv.2411.06648}\BibitemShut {NoStop}%
\bibitem [{\citenamefont {Garratt}\ and\ \citenamefont {Altman}(2024)}]{AltmanPRXQuantum5030311}%
  \BibitemOpen
  \bibfield  {author} {\bibinfo {author} {\bibfnamefont {S.~J.}\ \bibnamefont {Garratt}}\ and\ \bibinfo {author} {\bibfnamefont {E.}~\bibnamefont {Altman}},\ }\href {\doibase 10.1103/PRXQuantum.5.030311} {\bibfield  {journal} {\bibinfo  {journal} {PRX Quantum}\ }\textbf {\bibinfo {volume} {5}},\ \bibinfo {pages} {030311} (\bibinfo {year} {2024})}\BibitemShut {NoStop}%
\end{thebibliography}%
\let\addcontentsline\oldaddcontentsline
\onecolumngrid

\end{document}